\documentclass[preprint,12pt]{elsarticle}

\usepackage{graphicx}
\usepackage{amsmath,amssymb,bm}
\usepackage{booktabs,multirow}
\usepackage[linesnumbered,ruled,longend]{algorithm2e}
\usepackage{hyperref}
\usepackage{xcolor}


\journal{Computer Physics Communications}

\newcommand{\SF}{\mathrm{SF}}
\newcommand{\Dsel}{\mathcal D_{\mathrm{TBSCI}}}
\newcommand{\Dtpb}{\mathcal D_{\mathrm{TPB}}}

\newcommand{\draftfigure}[2]{%
\IfFileExists{#1}{%
\includegraphics[width=#2]{#1}%
}{%
\fbox{\parbox{0.85\linewidth}{%
Figure file unavailable:
\texttt{\detokenize{#1}}.}}%
}%
}

\begin{document}

\begin{frontmatter}

\title{Communication-Efficient Hamiltonian Evaluation for Massively Parallel Tensor-Product Bitstring Selected Configuration Interaction}

\author[ccs]{Enhua Xu\corref{cor1}}
\ead{enhua.xu@riken.jp}

\author[ithems,ccs]{Himadri Pathak}

\author[ccs]{William Dawson}

\author[ccs]{Takahito Nakajima}

\cortext[cor1]{Corresponding author}

\address[ccs]{RIKEN Center for Computational Science,
Kobe, Japan}

\address[ithems]{RIKEN Center for Interdisciplinary
Theoretical and Mathematical Sciences,
Wako, Japan}

\begin{abstract}
Extreme-scale selected configuration interaction (SCI) requires coordinated Hamiltonian evaluation, distributed CI-vector storage, and interprocess communication on massively parallel computers.
Our previous work [J. Chem. Phys. 164, 194103 (2026)] introduced the tensor-product bitstring selected configuration interaction (TBSCI) framework, combining on-the-fly Hamiltonian evaluation with fully distributed CI-vector storage and enabling parallel diagonalization of trillion-determinant spaces on up to 54,000 Fugaku nodes.
Here, we introduce FLIP, a new Hamiltonian evaluation algorithm within TBSCI that reorganizes the CI-vector layout and excitation-link data to replace remote CI-coefficient transfers for $\alpha$-electron double excitations with local coefficient access and link retrieval from nearby MPI ranks.
In communication-intensive full configuration interaction (FCI) benchmarks on Fugaku, no exposed communication waiting time was recorded on the slowest process across all 32 measured Hamiltonian--vector multiplications.
For N$_2$/cc-pVTZ with $2.62\times10^{12}$ determinants, the multiplication achieved 91.3\% strong-scaling efficiency from 18,000 to 81,000 nodes (more than 3.8 million compute cores).
We also obtained an FCI reference potential-energy curve for N$_2$/cc-pVTZ.
At all six bond lengths, spin bitstrings selected by their collective weights in reference SCI wavefunctions yielded variational TBSCI energies within 0.1~m$E_h$ of FCI using fewer than 1\% of the FCI determinants, demonstrating the compactness of the tensor-product bitstring representation.
Finally, we discuss local construction of double-excitation links to substantially reduce communication volume and eliminate persistent storage of precomputed double-link tables for future second-order perturbation theory calculations with large external determinant spaces.

\end{abstract}


\end{frontmatter}

\section{Introduction}
Full configuration interaction (FCI) provides an exact solution to the Schr\"odinger equation within a chosen finite orbital space, but the combinatorial growth of its determinant expansion severely limits its applicability.
Selected configuration interaction (SCI) offers a computationally more affordable alternative by retaining only the most important determinants in the configuration interaction (CI) expansion. 
Its variational accuracy can be systematically improved by including additional determinants, and the resulting energy is often supplemented by a perturbative correction.
Several SCI methods have been developed, including configuration interaction using a perturbative selection made iteratively (CIPSI),\cite{CIPSI,QUANTUM} adaptive sampling CI (ASCI),\cite{ASCI_2016,ASCI_2023} adaptive CI (ACI),\cite{ACI_2016} heat-bath CI (HCI),\cite{SHCI_2016,SHCI_2017,SHCI_2018} and iterative CI (iCI).\cite{iCI_2016,iCI_2020,iCI_2021} These methods have been successfully applied to strongly correlated electronic systems.

Extending SCI calculations to extreme-scale determinant spaces requires efficient use of computational and memory resources on massively parallel computers. 
Iterative eigensolvers such as the Davidson method\cite{Davidson1975} obtain the desired CI eigenstates through repeated Hamiltonian--vector multiplications, $\mathbf W=H\mathbf U$, where $H$ is the Hamiltonian matrix, $\mathbf U$ is a trial CI vector, and $\mathbf W$ is the resulting vector. 
The performance of this operation depends strongly on how the Hamiltonian and CI vectors are represented, stored, and accessed. 
The Hamiltonian matrix elements can either be stored explicitly or evaluated on the fly, while the CI vectors can be replicated or distributed across processes. Storing the sparse Hamiltonian avoids repeated evaluation of matrix elements but requires substantial memory. In contrast, evaluating matrix elements on the fly, commonly referred to as matrix-free Hamiltonian evaluation, reduces memory requirements but requires efficient identification of connected determinants. 
Similarly, replicating the CI vectors simplifies coefficient access but increases memory consumption on each process, whereas distributing them reduces per-process memory requirements but complicates coefficient access and introduces interprocess communication.
At such scales, parallel CI calculations therefore require a careful balance between computational cost, memory consumption, and communication overhead.

Different parallel SCI methods employ different strategies to address these challenges. 
The distributed-memory HCI implementation described in Ref.~\cite{SHCI_2018} distributes the sparse Hamiltonian across nodes while maintaining a full-length accumulation vector on each node during Hamiltonian--vector multiplication.
The distributed-memory ASCI implementation of Williams-Young et al.~\cite{ASCI_2023} distributes both the sparse Hamiltonian and CI vectors across processes, but still stores the Hamiltonian matrix elements explicitly.
Thus, fully distributed CI-vector storage has already been demonstrated in previous SCI implementations, although its combination with matrix-free Hamiltonian evaluation presents additional computational and communication challenges.
Quantum Package 2.0 evaluates Hamiltonian matrix elements on the fly while maintaining a copy of the input CI vector on each worker process.\cite{QUANTUM} 
The reported formal computational cost of identifying connected determinants scales as $\mathcal{O}(N_{\mathrm{det}}^{3/2})$, where $N_{\mathrm{det}}$ denotes the number of selected determinants. The authors noted that this cost could be reduced to $\mathcal{O}(N_{\mathrm{det}})$ by storing lists of single and double excitations for each spin, but this approach was not adopted in that implementation because of the additional memory requirements. 
On-the-fly Hamiltonian evaluation exploiting spin-string excitation connectivity was already established in direct FCI algorithms in the 1980s.\cite{FCI_1984_X,FCI_1984,FCI_1988}
Related string-based techniques were extended to SCI spaces in the 1990s.\cite{CPL_1992,TCA_1995}
Nevertheless, combining matrix-free Hamiltonian evaluation with fully distributed CI vectors had remained a major unresolved challenge for general SCI spaces, where irregular determinant selection complicates connectivity searches and interprocess data access.

In our previous work,\cite{TBSCI_202603} we introduced the tensor-product bitstring selected configuration interaction (TBSCI) framework and developed what is, to the best of our knowledge, the first SCI eigensolver to combine on-the-fly Hamiltonian evaluation with fully distributed CI-vector storage.
Each determinant is represented by an $\alpha$- and a $\beta$-bitstring.
Selected sets of these bitstrings define the tensor-product bitstring (TPB) determinant space, $D_{\mathrm{TPB}}$, comprising all products of the selected spin bitstrings.
The more general TBSCI determinant space, $D_{\mathrm{TBSCI}}\subseteq D_{\mathrm{TPB}}$, can retain an arbitrary subset of these determinants.
The TPB structure provides a unified framework for determinant indexing, Hamiltonian traversal, distributed CI-vector storage, and MPI communication, whether or not all products are retained.

The original TBSCI algorithm classifies $\alpha$-bitstring pairs by excitation degree and reuses precomputed single- and double-excitation links within the selected $\beta$-bitstring set, avoiding repeated connectivity searches during Hamiltonian evaluation.
The single-excitation links are replicated across processes, whereas the much larger double-excitation links are distributed to reduce memory requirements.
Excitation- and symmetry-based screening limit unnecessary data transfers, while excitation-aware bitstring ordering favors communication between nearby processes.
Asynchronous one-sided MPI operations further allow communication to overlap with computation.
In May 2025, TBSCI had demonstrated FCI calculations involving $2.62\times10^{12}$ determinants for N$_2$/cc-pVTZ.\cite{XuArxivHistory}
In March 2026, TBSCI executable binaries and representative benchmark data were made publicly available on GitHub.\cite{TBSCI_GITHUB_202603}
The work was subsequently published in May 2026, with the journal article reporting scaling to 54,000 Fugaku nodes (more than 2.5 million compute cores).\cite{TBSCI_202603}
We also demonstrated the structural compactness of the TPB representation: selecting spin bitstrings according to their collective weights in reference HCI wavefunctions\cite{SHCI_2018} yielded energies approaching the FCI limit while retaining only a small fraction of the FCI determinants.
Shirakawa {\it et al.} further demonstrated the applicability of a similar tensor-product determinant construction in a closed-loop quantum--classical workflow coupling Fugaku to IBM Heron quantum processors.
Their work, submitted to arXiv in October 2025,\cite{ShirakawaArxivHistory} employed multiple walkers to perform distributed diagonalization in subspaces containing up to $10^{10}$ determinants each, collectively using up to 152,064 Fugaku nodes;\cite{Shirakawa2026} further optimizations of the classical solver are being pursued by a growing number of researchers.~\cite{Nation2026Fulqrum}

An important motivation for further developing TBSCI is its extension to extreme-scale second-order perturbation theory (PT2) calculations. The external determinant space connected to a variational SCI wavefunction can exceed the variational space by orders of magnitude, placing greater demands on memory and interprocess communication.

In the original $\alpha$-based CI-vector layout, coefficients sharing the same $\alpha$-bitstring are grouped into segments and distributed across processes. 
Hamiltonian contributions are classified according to the numbers of $\alpha$- and $\beta$-electron excitations, denoted by $[a,b]$. The $[0,2]$ contribution is evaluated using local CI coefficients and distributed $\beta$-double-excitation links, whereas the $[2,0]$ contribution may require CI coefficients from remote processes.

Here, we introduce FLIP, a Hamiltonian evaluation algorithm within TBSCI that exploits complementary distributed CI-vector layouts to reorganize the two same-spin double-excitation contributions. 
FLIP retains the original $\alpha$-based layout for the $[0,2]$ contribution and constructs a complementary distributed $\beta$-based copy of the CI vector, allowing the $[2,0]$ contribution to be evaluated using local CI coefficients and separately retrieved $\alpha$-double-excitation links, analogous to the original $[0,2]$ evaluation.
These links are distributed in cyclically repeated groups across MPI processes to favor retrieval from nearby processes in the MPI rank ordering.
The link-based evaluation also permits local construction of double-excitation links from spin bitstrings, substantially reducing communication volume and eliminating the need for persistent double-link storage in future PT2 calculations.

We describe FLIP and analyze its computational and communication requirements for both FCI and general TBSCI spaces. We then evaluate its parallel performance on Fugaku using Hamiltonian--vector multiplications in FCI spaces as communication-intensive stress tests. 
Across all tested systems and node counts, no exposed communication waiting time was recorded on the slowest process during Hamiltonian--vector multiplication.
For N$_2$/cc-pVTZ, involving 2.62 trillion determinants, the Hamiltonian--vector multiplication achieves 91.3\% strong-scaling efficiency from 18,000 to 81,000 nodes (more than 3.8 million compute cores). We also obtain an FCI reference potential-energy curve for N$_2$/cc-pVTZ and demonstrate the structural compactness of the TPB representation: TBSCI energies differ from FCI by less than 0.1~m$E_h$ while retaining fewer than 1\% of the FCI determinants at all six sampled bond lengths. Finally, we discuss extensions toward PT2 calculations.

 \section{Methodology}
\label{sec:methods}

\subsection{Overview of the original TBSCI algorithm}
\label{sec:original}

In TBSCI, each determinant is represented by an $\alpha$-bitstring $S_w^\alpha$ and a $\beta$-bitstring $S_u^\beta$. Let the selected spin-bitstring sets $\{S^\alpha\}$ and $\{S^\beta\}$ contain $l_\alpha$ and $l_\beta$ elements, respectively. Their TPB determinant space comprises all products of the selected spin bitstrings,
\begin{equation}
\Dtpb=\bigl\{|S_w^\alpha\rangle\otimes|S_u^\beta\rangle:
S_w^\alpha\in\{S^\alpha\},\ S_u^\beta\in\{S^\beta\}\bigr\},
\qquad N_{\mathrm{TPB}}=l_\alpha l_\beta.
\label{eq:product}
\end{equation}

The more general TBSCI space, $\Dsel\subseteq\Dtpb$, can retain an arbitrary subset of these determinants. For each $S_w^\alpha$, define $\mathcal I_w^{(\beta)}$ as the set of its retained $\beta$ partners; analogously, $\mathcal J_u^{(\alpha)}$ contains the retained $\alpha$ partners of $S_u^\beta$. The same TBSCI wavefunction can then be written in either order:
\begin{equation}
\begin{split}
|\Psi_{\mathrm{TBSCI}}\rangle
&=\sum_{w=1}^{l_\alpha}\sum_{S_u^\beta\in\mathcal I_w^{(\beta)}}
c_{(w,u)}|S_w^\alpha\rangle\otimes|S_u^\beta\rangle \\
&=\sum_{u=1}^{l_\beta}\sum_{S_w^\alpha\in\mathcal J_u^{(\alpha)}}
c_{(w,u)}|S_w^\alpha\rangle\otimes|S_u^\beta\rangle.
\end{split}
\label{eq:wavefunction}
\end{equation}
The CI coefficients $c_{(w,u)}$ are independent variational parameters and are not assumed to factorize into separate $\alpha$- and $\beta$-dependent factors.
In the original $\alpha$-based layout, the $\beta$ index varies fastest. Coefficients sharing the same $\alpha$-bitstring form a segment, and these segments are distributed across MPI processes. 
Process $p$ stores $m_p$ segments and their retained partner lists. The local segments are numbered $i=1,\ldots,m_p$. The $\alpha$-bitstring of segment $i$ is denoted by $S_{p,i}^\alpha$.
Segment offsets and lengths define the locations of coefficient segments in the distributed vector, while sorted partner lists enable binary searches to locate
retained determinants and their corresponding coefficients.
In the complementary $\beta$-based layout, coefficients sharing the same $\beta$-bitstring form a segment, and these segments are likewise distributed across MPI processes.
The two layouts represent the same TBSCI wavefunction and differ only in the organization and distribution of its CI coefficients.

With distributed CI-vector storage, process $p$ computes its local portion of $\mathbf W=H\mathbf U$ as
\begin{equation}
\mathbf W_p=\sum_q H_{p,q}\mathbf U_q,
\label{eq:distributed_mv}
\end{equation}
where $H_{p,q}$ denotes the Hamiltonian block connecting the input coefficients owned by process $q$ to the output coefficients owned by process $p$. These blocks are not stored explicitly; their off-diagonal matrix elements are evaluated on the fly according to the Slater--Condon rules, while diagonal elements are precomputed and stored. As in our previous work,\cite{TBSCI_202603} a Hamiltonian contribution is classified as $[a,b]$, where $a$ and $b$ denote the numbers of $\alpha$- and $\beta$-electron excitations, respectively, between two determinants. For a Hamiltonian containing at most two-electron interactions, the six possible classes are $[0,0]$, $[1,0]$, $[0,1]$, $[2,0]$, $[1,1]$, and $[0,2]$.

Algorithm~\ref{alg:interaction_terms} summarizes the original Hamiltonian evaluation.\footnote{The membership test in the $[0,2]$ branch was inadvertently omitted from the pseudocode in our previous paper,\cite{TBSCI_202603} although it was already present in the implementation.}
In the $\alpha$-based layout, the $[2,0]$, $[1,0]$, and $[1,1]$ contributions involve coefficients in other $\alpha$ segments, which may reside on remote processes.
By contrast, $[0,2]$ and $[0,1]$ are evaluated within a local $\alpha$ segment.
The $[0,1]$, $[1,1]$, and $[0,2]$ contributions exploit precomputed $\beta$-excitation links.
The \texttt{BETA\_SINGLE\_LINK} and \texttt{BETA\_DOUBLE\_LINK} tables list, for each selected $\beta$-bitstring, the selected $\beta$-bitstrings connected to it by single and double excitations, respectively.
The smaller single-excitation table is replicated on every process, whereas the larger double-excitation table is divided into approximately equal batches, with one batch stored on each process.
Each link record identifies the connected bitstring and the orbital substitutions, together with the associated fermionic permutation sign needed to determine the Hamiltonian matrix element.

On Fugaku, the hybrid MPI/OpenMP implementation uses one MPI process per node and 48 computational cores. In each asynchronous multiplication step, one thread issues one-sided \texttt{MPI\_Get} requests for data needed in the next step, while the other 47 threads compute the current contribution in parallel. If $t_{\mathrm{data}}$ is the data-retrieval wall time and $t_{\mathrm{cal}}$ is the concurrently measured computation wall time, the exposed waiting time is
\begin{equation}
t_{\mathrm{delay}}=\max\{0,t_{\mathrm{data}}-t_{\mathrm{cal}}\},
\qquad T_{\mathrm{delay}}=\sum_{\mathrm{steps}}t_{\mathrm{delay}},
\label{eq:timing}
\end{equation}
where $t_{\mathrm{delay}}$ is the exposed communication waiting time in one asynchronous step, and $T_{\mathrm{delay}}$ sums it over all steps on a given process during one Hamiltonian--vector multiplication.
Thus, $T_{\mathrm{delay}}=0$ means that no exposed communication waiting was recorded; it does not imply the absence of data transfers or other communication-related overhead.

In the original TBSCI algorithm, CI-vector segments are ordered by the excitation degree of their $\alpha$-bitstrings relative to the Hartree--Fock (HF) $\alpha$-bitstring and assigned to processes in that order. A process retrieves a remote segment only if its $\alpha$-bitstring is connected to a local one by at most two excitations.\cite{TBSCI_202603} These strategies restrict data transfers and favor locality in the distributed layout. 
The $[2,0]$ contribution typically accesses substantially more CI-vector segments than the $[1,0]$ and $[1,1]$ contributions because double substitutions connect many more distinct $\alpha$-bitstrings. 
These connections also span a wider range of processes, increasing both the communication volume and the retrieval range in the MPI rank ordering.

\begin{algorithm}[!htbp]
\footnotesize\linespread{0.95}\selectfont
\DontPrintSemicolon
\SetAlgoNoLine
\SetArgSty{textup}
\KwData{Current process $p$, current step $q$; BETA\_SINGLE\_LINK and BETA\_DOUBLE\_LINK lists}
\For{$S_w^\alpha \leftarrow S^{\alpha}_{p,1}$ \KwTo $S^{\alpha}_{p,m_p}$}{
  \For{$S_x^\alpha \leftarrow S^{\alpha}_{q,1}$ \KwTo $S^{\alpha}_{q,m_q}$}{
    case $\gets$ \textbf{DIFF}($S_w^\alpha$, $S_x^\alpha$)\;
    \uIf{case $=2$}{
      \ForEach{$S_v^\beta \in \mathcal{I}_x^{(\beta)} $}{
        $S_u^\beta \gets S_v^\beta$\;
        If $S_u^\beta \in \mathcal{I}_w^{(\beta)}$, calculate $\langle S_{w}^\alpha | \langle S_{u}^\beta | \hat{H} | S_{x}^\alpha \rangle | S_{v}^\beta \rangle$ as [2,0] term\;
      }
    }
    \uElseIf{case $=1$}{
      \ForEach{$S_v^\beta \in \mathcal{I}_x^{(\beta)} $}{
        \ForEach{$S_u^\beta \in \text{BETA\_SINGLE\_LINK}(:,S_v^\beta)$}{
          If $S_u^\beta \in \mathcal{I}_w^{(\beta)}$, calculate $\langle S_{w}^\alpha | \langle S_{u}^\beta | \hat{H} | S_{x}^\alpha \rangle | S_{v}^\beta \rangle$ as [1,1] term\;
        }
        $S_u^\beta \gets S_v^\beta$\;
        If $S_u^\beta \in \mathcal{I}_w^{(\beta)}$, calculate $\langle S_{w}^\alpha | \langle S_{u}^\beta | \hat{H} | S_{x}^\alpha \rangle | S_{v}^\beta \rangle$ as [1,0] term\;
      }
    }
    \ElseIf{case $=0$}{
      \ForEach{$S_v^\beta \in \mathcal{I}_x^{(\beta)} $}{
        \ForEach{$S_u^\beta \in \text{BETA\_SINGLE\_LINK}(:,S_v^\beta)$}{
          If $S_u^\beta \in \mathcal{I}_w^{(\beta)}$, calculate $\langle S_{w}^\alpha | \langle S_{u}^\beta | \hat{H} | S_{x}^\alpha \rangle | S_{v}^\beta \rangle$ as [0,1] term\;
        }
        $S_u^\beta \gets S_v^\beta$\;
        If $S_u^\beta \in \mathcal{I}_w^{(\beta)}$, use the stored diagonal element as [0,0] term\;
      }
    }
  }
  $S_x^\alpha \leftarrow S_w^\alpha$\;
  \ForEach{$S_v^\beta \in \mathcal{I}_x^{(\beta)}$}{
    \If{$S_v^\beta$ belongs to the $q$-th batch of $\{S^\beta\}$}{
      \ForEach{$S_u^\beta \in \text{BETA\_DOUBLE\_LINK}(:,S_v^\beta)$}{
        If $S_u^\beta \in \mathcal{I}_w^{(\beta)}$, calculate $\langle S_w^\alpha | \langle S_u^\beta | \hat{H} | S_x^\alpha \rangle | S_v^\beta \rangle$ as $[0,2]$ term\;
    }
  }
}
}
\caption{Original Hamiltonian evaluation on process $p$ at step $q$. \textbf{DIFF} returns the number of electron substitutions between two same-spin bitstrings. Each matrix element is multiplied by its input coefficient and accumulated in the corresponding local output element.}
\label{alg:interaction_terms}
\end{algorithm}

\subsection{Computational cost and communication volume}
\label{sec:cost}

To estimate computational and communication costs, we distinguish selection of spin bitstrings from selection of determinants within their tensor-product space.
Let $L_\alpha$ and $L_\beta$ denote the numbers of spin bitstrings in the corresponding complete FCI string spaces, and define the spin-bitstring selection fractions $f_\alpha=l_\alpha/L_\alpha$ and $f_\beta=l_\beta/L_\beta$. The determinant-level sparsity factor $\SF=N_{\mathrm{TBSCI}}/(l_\alpha l_\beta)$ measures the fraction of determinants retained within $\Dtpb$. The average numbers of retained partners per $\alpha$ and $\beta$ segment are therefore $l_\beta\SF$ and $l_\alpha\SF$, respectively. Under this definition, symmetry restrictions also contribute to $\SF$; retaining every symmetry-allowed product does not generally imply $\SF=1$.

Table~\ref{tab:TBSCI-cost} compares the three dominant Hamiltonian contributions under the assumption of balanced workloads across $N$ nodes. Here, $N_{\mathrm{occ}}^\sigma$ and $N_{\mathrm{vir}}^\sigma$ denote the numbers of occupied and virtual orbitals, respectively, for spin $\sigma$. 
The estimates approximate the retained fraction of single- and double-excitation partners by $f_\sigma$ and use the average segment lengths $l_\beta\SF$ and $l_\alpha\SF$. These averages neglect variations in connectivity and segment length within the selected space.
In this model, the $[2,0]$ coefficient-transfer estimate contains an additional factor $N_{\mathrm{occ}}^\alpha N_{\mathrm{vir}}^\alpha$ relative to $[1,1]$. The $[2,0]$ contribution therefore involves more extensive coefficient retrieval over a wider range of processes. By contrast, the $[0,2]$ contribution uses local CI coefficients and retrieves distributed \texttt{BETA\_DOUBLE\_LINK} records. 
For fixed spin-bitstring sets, retaining fewer determinants within $\Dtpb$ decreases $\SF$ and the estimated coefficient-transfer volume, while leaving the precomputed link tables unchanged.

When $(N_{\mathrm{occ}}^\alpha)^2(N_{\mathrm{vir}}^\alpha)^2f_\alpha$ and $(N_{\mathrm{occ}}^\beta)^2(N_{\mathrm{vir}}^\beta)^2f_\beta$ are comparable, the estimates in Table~\ref{tab:TBSCI-cost} give an approximate $[2,0]$-to-$[0,2]$ communication-record ratio of $l_\alpha\SF:N$, reducing to $l_\alpha:N$ when $\SF\simeq1$.
In sufficiently sparse TBSCI spaces, the double-link traffic associated with $[0,2]$ can therefore exceed the CI-coefficient traffic associated with $[2,0]$. 
Nevertheless, link-based evaluation provides two ways to reduce communication overhead.
FLIP uses cyclically repeated groups of precomputed link records to restrict retrieval to a small range of processes.
With more memory per node, a complete link table can be distributed over fewer nodes, further narrowing this range.
A future extension is to construct double-excitation links locally from spin bitstrings, substantially reducing communication volume by replacing transfers of double-link records with transfers of much smaller spin-bitstring data, as discussed in Sec.~\ref{sec:discussion}.

\begin{table}[htbp]
\centering
\caption{Estimated computational cost and communication volume per node for the three dominant Hamiltonian terms during one Hamiltonian--vector multiplication of the original TBSCI algorithm. Balanced workloads over $N$ nodes, with one MPI process per node, are assumed. Constant and logarithmic search factors are omitted. Volumes count coefficient or link records, rather than bytes; Coefficient-transfer estimates do not account for reuse of fetched CI-vector segments across local segments. Outer bitstring-pair comparisons and link construction are excluded.}
\label{tab:TBSCI-cost}
\footnotesize
\setlength{\tabcolsep}{4pt}
\begin{tabular}{ccc}
\toprule
Dominant term & Computational cost & Communication volume \\
\midrule
$[2,0]$ 
& $\displaystyle \frac{l_\alpha}{N} \left[ \left(N_{\mathrm{occ}}^\alpha\right)^2 \left(N_{\mathrm{vir}}^\alpha\right)^2 f_\alpha
\right] \left(l_\beta\mathrm{SF}\right)$ 
& $\displaystyle \frac{l_\alpha}{N} \left[ \left(N_{\mathrm{occ}}^\alpha\right)^2 \left(N_{\mathrm{vir}}^\alpha\right)^2 f_\alpha \right] \left(l_\beta\mathrm{SF}\right)$ \\
$[1,1]$
& $\displaystyle \frac{l_\alpha}{N} \left( N_{\mathrm{occ}}^\alpha N_{\mathrm{vir}}^\alpha f_\alpha \right) \left(l_\beta\mathrm{SF}\right) \left( N_{\mathrm{occ}}^\beta N_{\mathrm{vir}}^\beta f_\beta \right)$
& $\displaystyle \frac{l_\alpha}{N} \left( N_{\mathrm{occ}}^\alpha N_{\mathrm{vir}}^\alpha f_\alpha \right)\left(l_\beta\mathrm{SF}\right)$ \\
$[0,2]$
& $\displaystyle \frac{l_\alpha}{N} \left(l_\beta\mathrm{SF}\right) \left[ \left(N_{\mathrm{occ}}^\beta\right)^2 \left(N_{\mathrm{vir}}^\beta\right)^2 f_\beta \right]$
& $\displaystyle l_\beta \left[ \left( N_{\mathrm{occ}}^\beta \right)^2 \left( N_{\mathrm{vir}}^\beta \right)^2 f_\beta \right]$ \\
\bottomrule
\end{tabular}
\end{table}

\subsection{The FLIP algorithm}
\label{sec:flip}

\begin{figure*}[htbp]
\centering
\draftfigure{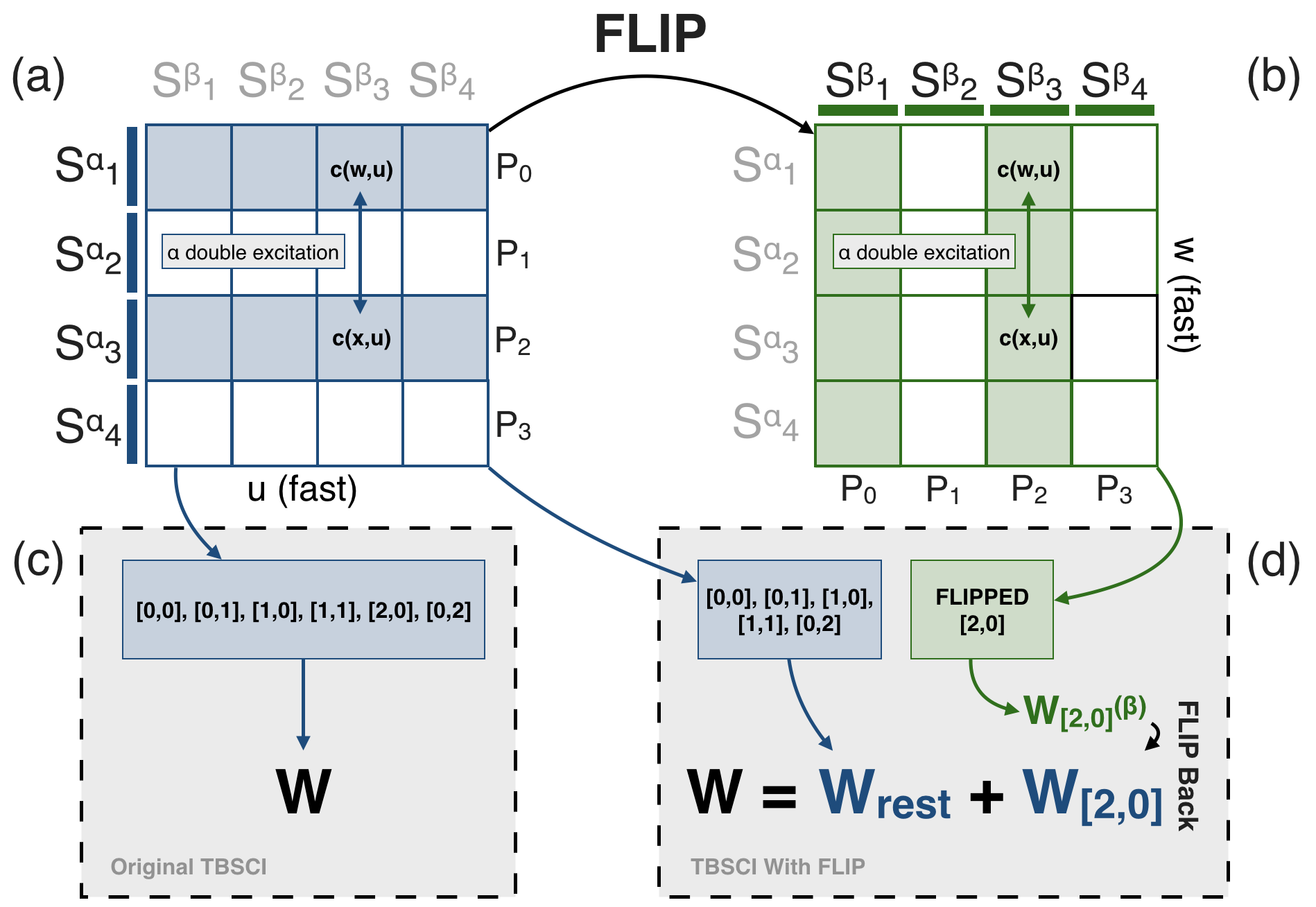}{\textwidth}
\caption{Schematic illustration of the FLIP algorithm and its complementary CI-vector layouts. Four MPI processes, each holding one coefficient segment, are shown for simplicity; in practice, each process holds many segments.
(a) In the original $\alpha$-based layout, the $\beta$-bitstring index $u$ varies fastest. The coefficients $c(w,u)$ and $c(x,u)$, connected by an $\alpha$-electron double excitation, reside on different
processes, requiring remote coefficient access for the $[2,0]$ contribution.
(b) FLIP redistributes a copy of the CI vector into the $\beta$-based layout, where the $\alpha$-bitstring index $w$ varies fastest and the coefficients $c(w,u)$ and $c(x,u)$ reside on the same process. The $[2,0]$ contribution can then be evaluated using local CI coefficients and distributed $\alpha$-double-excitation links.
(c) The original TBSCI algorithm evaluates all six Hamiltonian contributions in the $\alpha$-based layout.
(d) FLIP evaluates five contributions in the original layout and the $[2,0]$ contribution in the $\beta$-based layout. The latter partial output is transformed back to the $\alpha$-based layout before the two outputs are combined.}
\label{fig:diagram}
\end{figure*}

The same-spin double-excitation contributions can be evaluated using different data-access patterns under $\alpha$- and $\beta$-based CI-vector layouts \textcolor{blue}{(Fig.~\ref{fig:diagram})}. In the present implementation, FLIP retains the original $\alpha$-based distributed CI vector and creates a second, $\beta$-based distributed copy. Both layouts coexist during Hamiltonian--vector multiplication: 
the $\beta$-based copy evaluates $[2,0]$ using local CI coefficients and separately retrieved \texttt{ALPHA\_DOUBLE\_LINK} records, analogous to the $[0,2]$ evaluation in the original layout; the original $\alpha$-based layout evaluates the remaining five Hamiltonian contributions.
The two partial results are then combined in the original $\alpha$-based layout. The physical excitation remains $[2,0]$; only its data-access pattern changes.

To construct the $\beta$-based copy, each process uses precomputed index mappings to pack its coefficients into destination-process blocks. The blocks are exchanged by \texttt{MPI\_Alltoallv} and unpacked into the local $\beta$ segments. The mappings are constructed once and reused in subsequent iterations, while the original $\alpha$-based vector remains available.
After evaluating $[2,0]$ in the $\beta$-based layout, FLIP redistributes the resulting partial vector to the original layout using the reverse mapping and adds it to the other five contributions. 
Thus, the two layout transformations and the additional distributed copy do not change the Hamiltonian, the selected determinant space, or the CI coefficients.

Algorithm~\ref{alg:FLIP_term} evaluates the $[2,0]$ contribution in the $\beta$-based CI-vector layout, hereafter referred to as the flipped $[2,0]$ contribution. Using the same connectivity approximations as in Table~\ref{tab:TBSCI-cost} and assuming balanced workloads across $N$ nodes, we estimate its \textbf{C}omputational cost per node, $C_{[2,0]}^{\mathrm{FLIP}}$, and the \textbf{V}olume of link records retrieved per node, $V_{[2,0],\mathrm{link}}^{\mathrm{FLIP}}$, as
\begin{equation}
\begin{split}
C_{[2,0]}^{\mathrm{FLIP}}
&\sim \frac{l_\beta}{N}\,(l_\alpha\SF)\,
(N_{\mathrm{occ}}^\alpha)^2
(N_{\mathrm{vir}}^\alpha)^2 f_\alpha,\\
V_{[2,0],\mathrm{link}}^{\mathrm{FLIP}}
&\sim l_\alpha\,
(N_{\mathrm{occ}}^\alpha)^2
(N_{\mathrm{vir}}^\alpha)^2 f_\alpha.
\end{split}
\label{eq:flip_cost}
\end{equation}
The link-record communication estimate accounts for reuse across local $\beta$ segments and excludes communication associated with the two CI-vector layout transformations.
Comparing the estimates in Table~\ref{tab:TBSCI-cost} and Eq.~\eqref{eq:flip_cost} gives an original-to-FLIP communication-record ratio of $l_\beta\SF:N$ for the $[2,0]$ contribution.
In the present implementation, each CI coefficient occupies 8 bytes, while each double-excitation link record contains a 4-byte bitstring index, four 1-byte orbital indices, and a 1-byte sign factor, totaling 9 bytes.
The orbital indices can be safely stored in one byte for the active spaces considered here.
The corresponding estimated communication-volume ratio in bytes is therefore $8l_\beta\SF:9N$.

The leading computational scaling is unchanged, but the flipped loop ordering introduces repeated accesses to the integral and permutation-sign information.
For a fixed pair of $\alpha$-bitstrings connected by a double excitation, the corresponding Hamiltonian matrix element is an antisymmetrized two-electron integral multiplied by the fermionic permutation sign and is independent of the common $\beta$-bitstring.
Algorithm~\ref{alg:interaction_terms} can obtain this matrix element once before traversing the retained $\beta$ partners. 
In Algorithm~\ref{alg:FLIP_term}, the integral and permutation-sign information is accessed within the innermost link traversal, and these accesses are repeated across $\beta$ segments, adding indexing and memory-access overhead.

For open-shell systems, we choose the spin labels so that the $\alpha$-bitstring set is the larger one.
In the CN/cc-pVTZ benchmark, $L_\alpha=4{,}582{,}116$ and $L_\beta=424{,}270$, and the $\alpha$ double-link table is also larger. 
The additional $\alpha$-link storage and the increased computational overhead are particularly relevant to this system, as discussed below.

Cyclic link storage further restricts the range of processes accessed during link retrieval.
In the original TBSCI algorithm, a complete double-link table is divided into $N$ batches, with one batch stored on each process.
FLIP organizes these batches into $N_{\mathrm{groups}}$ groups and replicates the groups cyclically across processes, with one group stored on each process.
The same $N_{\mathrm{groups}}$ is used for the $\alpha$ and $\beta$ double-link tables.
Any $N_{\mathrm{groups}}$ consecutive processes therefore collectively hold a complete copy of each table, allowing each process to retrieve the required links from nearby processes in the MPI rank ordering.
In FLIP, the $[0,2]$ contribution retains the evaluation procedure of Algorithm~\ref{alg:interaction_terms}, but retrieves $\beta$-double-excitation links from cyclically replicated groups rather than the original batches.
Reducing $N_{\mathrm{groups}}$ narrows the retrieval range at the cost of increased per-process link storage.
In practice, $N_{\mathrm{groups}}$ is chosen to be of the order of $N_{\mathrm{occ}}N_{\mathrm{vir}}$, balancing communication locality against per-process double-link storage.

The $\beta$-based FLIP operation uses two additional distributed double-precision arrays and two distributed 4-byte integer mapping arrays.
For N$_2$/cc-pVTZ, these arrays require approximately 3.3 GiB per node at the smallest tested allocation of 18,000 nodes.
Each of the $\alpha$- and $\beta$-double-excitation link tables contains approximately $7.94\times10^9$ records, so one complete copy of both tables occupies approximately 133~GiB.
Their per-node storage under cyclic replication depends on $N_{\mathrm{groups}}$, which varies with the system and node allocation.
For this calculation, the two link tables together occupy less than 0.5~GiB per node on average.

\begin{algorithm}[!ht]
\DontPrintSemicolon
\SetAlgoNoLine
\SetArgSty{textup}
\KwData{Current process $p$, current step $q$; ALPHA\_DOUBLE\_LINK list}
\For{$S_u^\beta \leftarrow S^{\beta}_{p,1}$ \KwTo $S^{\beta}_{p,n_p}$}{
  $S_v^\beta \leftarrow S_u^\beta$\;
  \ForEach{$S_x^\alpha \in \mathcal{J}_v^{(\alpha)}$}{
  \If{the double-excitation links for $S_x^\alpha$ belong to a group processed by $p$ at step $q$}{
      \ForEach{$S_w^\alpha \in \text{ALPHA\_DOUBLE\_LINK}(:,S_x^\alpha)$}{
        If $S_w^\alpha \in \mathcal{J}_u^{(\alpha)}$, calculate $\langle S_w^\alpha | \langle S_u^\beta | \hat{H} | S_x^\alpha \rangle | S_v^\beta \rangle$ as flipped $[2,0]$ term\;
      }
    }
  }  
}
\caption{Flipped $[2,0]$ evaluation on process $p$ at step $q$ in the $\beta$-based layout, where $n_p$ is the number of local $\beta$ segments. Each matrix element is multiplied by its local input coefficient and accumulated in the local partial output vector.}
\label{alg:FLIP_term}
\end{algorithm}

\subsection{Computational setup}

All TBSCI calculations were performed on Fugaku using double-precision arithmetic for CI coefficients and Hamiltonian matrix elements.
HF orbitals and molecular integrals were generated using PySCF~\cite{PYSCF} and supplied to TBSCI through FCIDUMP files.
As in our previous study~\cite{TBSCI_202603}, all calculations reported here retained every symmetry-allowed determinant within $\Dtpb$.
More generally, FLIP can be applied to arbitrary selected subsets of $\Dtpb$.

Fugaku's Tofu Interconnect D provides system-specific MPI settings that affect communication performance and stability.
In our previous study,~\cite{TBSCI_202603} four Tofu-related MPI MCA parameters were explicitly configured to maximize communication-related resources, while \texttt{UTOFU\_SWAP\_PROTECT} remained at its default value of 0, allowing communication memory regions to be swapped out.
To improve communication stability without excessive memory consumption, we removed these explicit MCA settings, allowing the system to use their default values, and enabled \texttt{UTOFU\_SWAP\_PROTECT=1} to prevent communication memory regions from being swapped out.
Both the original TBSCI algorithm and FLIP were benchmarked under this revised configuration, and all benchmark jobs completed successfully.
The Hamiltonian--vector multiplication times of FLIP were nearly unchanged by these configuration changes.
For the original algorithm, their effect on performance could not be reliably determined because $T_{\mathrm{delay}}$ varied considerably between independent job submissions.

\section{Results}
\label{sec:results}

\subsection{Strong scaling on Fugaku}
\label{sec:scaling}

We assess FLIP using the four FCI systems benchmarked in our previous TBSCI study.\cite{TBSCI_202603} Their FCI dimensions and the node ranges used here are given in Table~\ref{tab:systems}. The complete spin-bitstring sets are retained, with determinants restricted to the specified molecular symmetry. For the systems considered here, this gives $\SF\simeq1/4$ for $C_{2v}$ and $\SF\simeq1/8$ for $D_{2h}$. 
Approximately equal numbers of CI coefficients were assigned to the MPI processes.

\begin{figure}[htbp]
\centering
\draftfigure{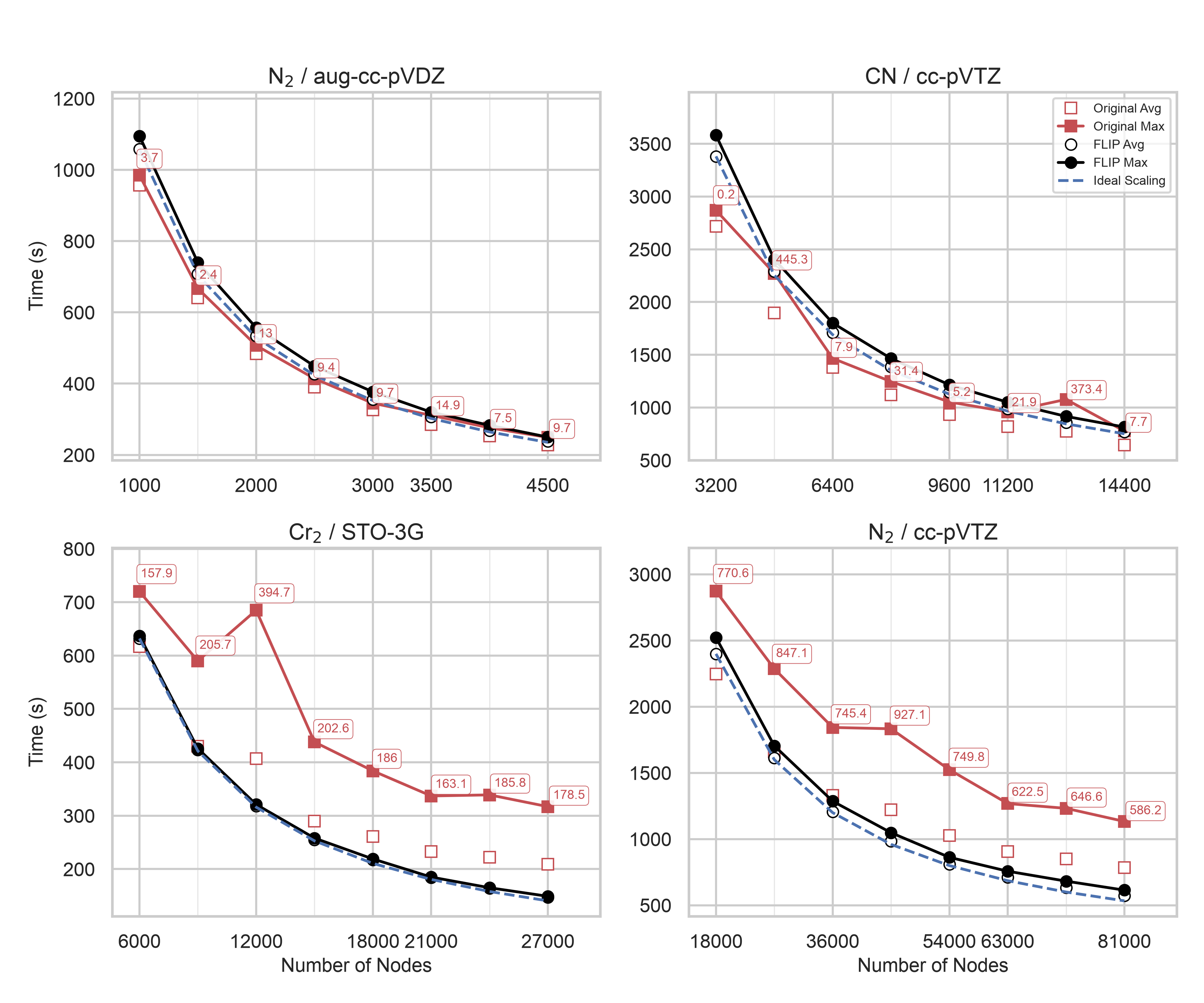}{\textwidth}
\caption{Strong scaling of Hamiltonian--vector multiplication on Fugaku. Red squares indicate the original TBSCI algorithm and black circles FLIP; open and filled symbols show average and maximum times across nodes, respectively. Boxed values give $T_{\mathrm{delay}}$ on the slowest process, in seconds. 
Dashed lines show ideal inverse-node-count scaling anchored at the FLIP $T_{\mathrm{avg}}$ value at the smallest node count; the reported strong-scaling efficiencies are calculated from $T_{\mathrm{max}}$.}
\label{fig:timing}
\end{figure}

Our previous TBSCI study reported nearly constant per-iteration wall times throughout the diagonalization procedure.\cite{TBSCI_202603}
Similar stability was observed across iterations in the present FLIP calculations.
To limit computational cost, we report the first recorded Hamiltonian--vector multiplication for each allocation, without averaging over repeated runs. 
The Hamiltonian--vector multiplication timings exclude the two CI-vector layout transformations. 
Figure~\ref{fig:timing} shows the mean time across nodes, $T_{\mathrm{avg}}$, and the longest time, $T_{\mathrm{max}}$, which determines completion of the distributed multiplication. The figure also reports the exposed communication waiting time, $T_{\mathrm{delay}}$, on the slowest process, as defined in Eq.~\eqref{eq:timing}. 
The individual timings are listed in Tables~S1--S4 of the Supplementary Material. Strong-scaling efficiency is calculated from the maximum times at the smallest and largest node counts, $N_0$ and $N_1$, respectively, for each benchmark system:
\begin{equation}
\eta_{\mathrm{FLIP}}=
\frac{T_{\mathrm{max}}(N_0)N_0}
{T_{\mathrm{max}}(N_1)N_1}.
\label{eq:efficiency}
\end{equation}

Both implementations use precomputed double-excitation links.
For the $[2,0]$ contribution, the original algorithm retrieves doubly connected CI-vector segments, whereas FLIP retrieves \texttt{ALPHA\_DOUBLE\_LINK} records and uses local CI coefficients.
For the FCI spaces considered here, the cost estimates in Secs.~\ref{sec:cost} and \ref{sec:flip} suggest that the link-based pathway can reduce communication requirements relative to the original coefficient-based pathway.
Moreover, the cyclic replication of link-data groups allows each process to retrieve the required links from a window of $N_{\mathrm{groups}}$ consecutive MPI ranks rather than from the full allocation.
These potential savings compete with the additional local computation of the flipped $[2,0]$ traversal, and their combined effect is reflected in the measured $T_{\mathrm{avg}}$, $T_{\mathrm{max}}$, and $T_{\mathrm{delay}}$.

For the slowest process in each allocation,  FLIP records no exposed communication waiting ($T_{\mathrm{delay}}=0$) in any of the 32 measured multiplications across the four systems, including the largest calculation on 81,000 nodes (more than 3.8 million compute cores). 
All four systems retain more than 91\% strong-scaling efficiency over a 4.5-fold increase in node count, with 91.3\% efficiency for N$_2$/cc-pVTZ with $2.62\times10^{12}$ determinants. At 81,000 nodes, the maximum multiplication time is 614~s with FLIP, compared with 1133~s and $T_{\mathrm{delay}}=586.2$~s for the original algorithm. For Cr$_2$/STO-3G at 27,000 nodes, FLIP similarly reduces the maximum time from 317 to 149~s, while the original algorithm records 178.5~s of exposed communication waiting.
Zero exposed communication waiting means that data retrieval completed within the overlapping computation on the slowest process; however, load imbalance and incomplete utilization of computational resources can still limit strong-scaling efficiency.

When exposed communication waiting is already small, the additional computation in the flipped $[2,0]$ traversal may offset its benefit. For N$_2$/aug-cc-pVDZ, the maximum times of the two algorithms are nearly identical at 4,500 nodes (249~s for the original algorithm and 250~s for FLIP). For the open-shell CN/cc-pVTZ system at 3,200 nodes, where the original $T_{\mathrm{delay}}$ is only 0.2~s, FLIP increases the maximum time from 2870 to 3582~s. Conversely, at 12,800 nodes, where the original waiting time reaches 373.4~s, FLIP reduces the maximum time from 1077 to 916~s. 
These results show that FLIP becomes advantageous when the reduction in exposed communication waiting outweighs the additional computational overhead of the flipped $[2,0]$ traversal.

Tables~S5--S8 of the Supplementary Material report wall times for a single Davidson iteration, including operations beyond Hamiltonian--vector multiplication.
For FLIP, the recorded times exceed the maximum Hamiltonian--vector multiplication times by 2--15~s (at most 2.1\%), including both CI-vector layout transformations and other Davidson operations.
The endpoint strong-scaling efficiencies based on these wall times remain above 91\% for all four systems with FLIP.

\begin{table}[htbp]
\centering
\caption{FCI benchmarks and strong-scaling results for FLIP. 
Two orbitals are frozen for both N$_2$ systems and CN, and 12 for Cr$_2$. 
The calculations use $C_{2v}$ symmetry for CN and $D_{2h}$ symmetry for the other systems.
$T_{\mathrm{max}}$ (s) is the longest time across nodes for one Hamiltonian--vector multiplication; the two $T_{\mathrm{max}}$ columns correspond to the first and last node counts in the range.}
\label{tab:systems}
\footnotesize
\setlength{\tabcolsep}{3pt}
\begin{tabular}{lcccccc}
\toprule
System & Active space & FCI dimension & Node range & \multicolumn{2}{c}{$T_{\mathrm{max}}$ (s)} & $\eta_{\mathrm{FLIP}}$ \\
\cmidrule(lr){5-6}
 & & & $N_0$--$N_1$ & $N_0$ & $N_1$ & \\
\midrule
N$_2$/aug-cc-pVDZ & CAS(10e,44o) & $1.47\times10^{11}$ & 1,000--4,500 & 1095 & 250 & 97.3\% \\
CN/cc-pVTZ & CAS(9e,58o) & $4.86\times10^{11}$ & 3,200--14,400 & 3582 & 815 & 97.7\% \\
Cr$_2$/STO-3G & CAS(24e,24o) & $9.14\times10^{11}$ & 6,000--27,000 & 637 & 149 & 95.0\% \\
N$_2$/cc-pVTZ & CAS(10e,58o) & $2.62\times10^{12}$ & 18,000--81,000 & 2523 & 614 & 91.3\% \\
\bottomrule
\end{tabular}
\end{table}

\subsection{FCI reference curve and TPB compactness for \texorpdfstring{N$_2$}{N2}/cc-pVTZ}
\label{sec:pes}

We obtained an FCI reference potential-energy curve for N$_2$/cc-pVTZ at six bond lengths ranging from 2.068 to 4.2~bohr and used these energies to assess the structural compactness of the TPB representation from equilibrium to stretched geometries.
At $R=2.068$, 2.4, and 2.7~bohr, we retained the FCI energies previously obtained with the original TBSCI algorithm under an energy-only convergence criterion: the absolute difference between successive Davidson energies was required to remain below $10^{-7}~E_h$ for two consecutive iterations.
For the more challenging stretched geometries at $R=3.0$, 3.6, and 4.2~bohr, we employed FLIP with the same energy criterion and an additional residual-norm threshold of $10^{-4}$.
For these three FLIP calculations, no exposed communication waiting was recorded on the slowest process during any Hamiltonian--vector multiplication.
This communication efficiency was particularly important at $R=4.2$~bohr, where the calculation converged in 94 Davidson iterations on 55,296 Fugaku nodes, taking 21.6 hours and thus finishing within Fugaku's 24-hour job limit.

To construct the selected spin-bitstring sets for the TPB compactness analysis, we generated an HCI reference wavefunction at each geometry using the DICE program with a selection threshold of $\theta=10^{-5}$.\cite{SHCI_2016,SHCI_2017}
For each normalized HCI reference wavefunction, we assigned a collective weight to each $\alpha$- or $\beta$-bitstring by summing the squared coefficients of all determinants containing that bitstring.
A spin bitstring was retained if its weight exceeded $\delta$ times the weight of the corresponding HF spin bitstring in the same reference wavefunction. For each threshold $\delta$, all symmetry-allowed products of the retained $\alpha$- and $\beta$-bitstrings were included in $\Dsel$. The Hamiltonian was then diagonalized in this space to obtain the variational TBSCI energy and wavefunction. At $\delta=0$, all spin bitstrings with nonzero weight in the HCI reference wavefunction were retained.

Table~\ref{tab:N2_TZ_PES} shows the rate at which the TBSCI energies approach the FCI reference as $\delta$ decreases at every geometry. At $\delta=10^{-9}$, the energy deviations range from 0.490 to 0.880~m$E_h$, while the retained spaces contain only 0.011--0.071\% of the FCI determinants. Retaining all spin bitstrings present in the HCI reference wavefunctions ($\delta=0$) reduces the deviations to 0.074--0.089~m$E_h$, with only 0.48--0.82\% of the FCI determinants. 
At $\delta=0$, the retained space increases from $1.26\times10^{10}$ determinants near equilibrium to $2.15\times10^{10}$ at 4.2~bohr, yet the energy deviations remain below 0.1~m$E_h$ at all six sampled bond lengths.

To further examine the effect of expanding the spin-bitstring sets, we present in Fig.~\ref{fig:histograms} the distributions of relative CI-coefficient magnitudes, $|c_K/c_{\mathrm{HF}}|$, at different selection thresholds. Here, $c_K$ and $c_{\mathrm{HF}}$ denote the coefficients of determinant $K$ and the HF determinant, respectively, in the same normalized TBSCI wavefunction. The histograms show the numbers of determinants within different ranges of relative coefficient magnitude on a logarithmic count scale.

At every bond length, the populations of the largest-coefficient bins change little as $\delta$ decreases, whereas those of the smaller-coefficient bins increase substantially.
Thus, expanding the spin-bitstring sets primarily introduces additional determinants with relatively small coefficients.
Together with the systematic convergence of the variational energies, these results demonstrate the structural compactness of the TPB representation for N$_2$/cc-pVTZ: selected spin-bitstring sets generate determinant spaces containing only a small fraction of the FCI determinants while yielding near-FCI energies across the sampled bond lengths.

\begin{table}[!htbp]
\centering
\caption{Variational TBSCI results for N$_2$/cc-pVTZ along the potential-energy curve. The relative bitstring-weight threshold $\delta$ determines the selected spin-bitstring sets of size $l_\alpha=l_\beta$; $N_{\mathrm{TBSCI}}$ is the number of their symmetry-allowed products. The FCI space contains $2{,}624{,}478{,}142{,}680$ determinants in the target symmetry sector of $D_{2h}$ with two frozen orbitals. Bond lengths $R$ are in bohr, HF and FCI energies are in hartree, and energy deviations $E_{\mathrm{TBSCI}}-E_{\mathrm{FCI}}$ are in millihartree (m$E_h$).}
\label{tab:N2_TZ_PES}
\begingroup
\footnotesize
\renewcommand{\arraystretch}{0.88}
\setlength{\aboverulesep}{0.25ex}
\setlength{\belowrulesep}{0.35ex}
\begin{tabular}{ccccc}
\toprule
$\delta$ & $l_\alpha=l_\beta$ & $N_{\mathrm{TBSCI}}$ & $N_{\mathrm{TBSCI}}/N_{\mathrm{FCI}}$ & $E_{\mathrm{TBSCI}} - E_{\mathrm{FCI}}$ (m$E_h$) \\
\midrule
\multicolumn{5}{c}{$R = 2.068$ bohr \hspace{0.6cm} $E_{\mathrm{HF}}=-108.984093$ \hspace{0.6cm} $E_{\mathrm{FCI}}=-109.375153$} \\
\midrule
$1\times10^{-6}$ & 1,530    & 586,634       & 0.000022\%   & 20.668  \\
$1\times10^{-7}$ & 5,472    & 5,147,844     & 0.00020\%    & 8.036 \\
$1\times10^{-8}$ & 16,861   & 42,751,597    & 0.0016\%     & 2.633 \\
$1\times10^{-9}$ & 43,914   & 288,849,332   & 0.011\%      & 0.844 \\
0                & 260,795  & 12,554,326,295 & 0.48\%      & 0.087 \\
\midrule
\multicolumn{5}{c}{$R = 2.4$ bohr \hspace{0.6cm} $E_{\mathrm{HF}}=-108.891508$ \hspace{0.6cm} $E_{\mathrm{FCI}}=-109.327963$} \\
\midrule
$1\times10^{-6}$ & 1,827   & 732,087       & 0.000028\% & 21.492 \\
$1\times10^{-7}$ & 6,308   & 6,654,932     & 0.00025\%  & 8.068 \\
$1\times10^{-8}$ & 18,787  & 52,778,125    & 0.0020\%   & 2.656 \\
$1\times10^{-9}$ & 48,157  & 341,787,783   & 0.013\%    & 0.865 \\
0                & 277,552 & 13,339,614,310 & 0.51\%    & 0.087 \\
\midrule
\multicolumn{5}{c}{$R = 2.7$ bohr \hspace{0.6cm} $E_{\mathrm{HF}}=-108.762233$ \hspace{0.6cm} $E_{\mathrm{FCI}}=-109.247223$} \\
\midrule
$1\times10^{-6}$ & 2,202   & 969,884     & 0.000037\%    & 21.795 \\
$1\times10^{-7}$ & 7,222   & 8,508,256     & 0.00032\%   & 8.130 \\
$1\times10^{-8}$ & 21,076  & 65,923,154    & 0.0025\%    & 2.726 \\
$1\times10^{-9}$ & 53,430  & 414,994,178   & 0.016\%     & 0.880 \\
0                & 295,908 & 14,315,646,382 & 0.55\%     & 0.089 \\
\midrule
\multicolumn{5}{c}{$R = 3.0$ bohr \hspace{0.6cm} $E_{\mathrm{HF}}=-108.631934$ \hspace{0.6cm} $E_{\mathrm{FCI}}=-109.171355$} \\
\midrule
$1\times10^{-6}$ & 2,591   & 1,227,493     & 0.000047\%  & 22.130 \\
$1\times10^{-7}$ & 8,342   & 11,127,010    & 0.00042\%   & 8.075 \\
$1\times10^{-8}$ & 24,272  & 85,142,246    & 0.0032\%    & 2.667 \\
$1\times10^{-9}$ & 60,113  & 512,048,041   & 0.020\%     & 0.858 \\
0                & 316,348 & 15,345,570,270 & 0.58\%     & 0.086 \\
\midrule
\multicolumn{5}{c}{$R = 3.6$ bohr \hspace{0.6cm} $E_{\mathrm{HF}}=-108.411469$ \hspace{0.6cm} $E_{\mathrm{FCI}}=-109.074400$} \\
\midrule
$1\times10^{-6}$ & 4,205   & 2,809,563     & 0.00011\%   & 19.562 \\
$1\times10^{-7}$ & 12,694  & 23,546,968    & 0.00090\%   & 7.216 \\
$1\times10^{-8}$ & 36,196  & 178,275,718   & 0.0068\%    & 2.258 \\
$1\times10^{-9}$ & 85,841  & 986,815,425   & 0.038\%     & 0.696 \\
0                & 373,725 & 19,200,449,783 & 0.73\%     & 0.075 \\
\midrule
\multicolumn{5}{c}{$R = 4.2$ bohr \hspace{0.6cm} $E_{\mathrm{HF}}=-108.250458$ \hspace{0.6cm} $E_{\mathrm{FCI}}=-109.041506$} \\
\midrule
$1\times10^{-6}$ & 6,516   & 6,147,156     & 0.00023\%   & 15.281 \\
$1\times10^{-7}$ & 19,115  & 49,983,061    & 0.0019\%    & 5.550 \\
$1\times10^{-8}$ & 52,874  & 366,187,572   & 0.014\%     & 1.634 \\
$1\times10^{-9}$ & 119,501 & 1,859,460,047 & 0.071\%     & 0.490 \\
0                & 404,000 & 21,532,534,112& 0.82\%     & 0.074 \\
\bottomrule
\end{tabular}
\endgroup
\end{table}

\begin{figure}[htbp]
\centering
\draftfigure{CI_histograms_N2.png}{\textwidth}
\caption{Distributions of relative CI-coefficient magnitudes for N$_2$/cc-pVTZ at six bond lengths $R$ (bohr). The five series correspond to bitstring-weight thresholds $\delta=10^{-6}$, $10^{-7}$, $10^{-8}$, $10^{-9}$, and $0$, with the last shown in red. The horizontal labels indicate coefficient-magnitude bins in $|c_K/c_{\mathrm{HF}}|$; the ordinate is the number of determinants. All series represent TBSCI wavefunctions.}
\label{fig:histograms}
\end{figure}

\section{Future extensions}
\label{sec:discussion}

In the present implementation, FLIP evaluates the $[2,0]$ contribution using local CI coefficients and precomputed double-excitation links retrieved from nearby processes in the MPI rank ordering.
As a future extension, each process could retrieve the required spin bitstrings and construct double-excitation links locally for reuse across its local CI-vector segments.
This would replace transfers of large volumes of double-link records with transfers of much smaller spin-bitstring data, substantially reducing communication volume.
Generating links as needed and discarding them after use would also eliminate the need for persistent storage of precomputed double-link tables.

This approach is particularly relevant to future PT2 calculations, where the external determinant space connected to the variational wavefunction can exceed the variational space by orders of magnitude, making the reduction in link-data communication especially valuable.
Although local link construction introduces additional computation, high computational throughput relative to network bandwidth may make this approach particularly attractive on GPU-based systems.
The overall runtime benefit will depend on the balance between the additional local computation and the reduced communication.

The complementary layouts could also allow the $[0,2]$ contribution to be evaluated using the original coefficient-based strategy for $[2,0]$, with the roles of $\alpha$ and $\beta$ exchanged.
In the $\beta$-based layout, this alternative retrieves CI coefficients from remote processes instead of double-excitation links. 
It should be advantageous in sufficiently sparse TBSCI spaces, where retrieving the required coefficients costs less than retrieving the corresponding link records. 
FLIP therefore provides flexibility to choose between coefficient-based and link-based evaluation of each same-spin double-excitation contribution according to the structure of the determinant spaces involved and the associated communication costs.

\section{Conclusions}
\label{sec:conclusions}

We have developed FLIP, a communication-efficient Hamiltonian evaluation algorithm within the TBSCI framework.
By retaining the original $\alpha$-based distributed CI vector and constructing a $\beta$-based copy, FLIP enables both same-spin double-excitation contributions, $[2,0]$ and $[0,2]$, to be evaluated using local CI coefficients.
For $[2,0]$, remote coefficient transfers are replaced by separately retrieved $\alpha$-double-excitation links, organized into cyclically repeated groups to favor communication between nearby processes in the MPI rank ordering.
The Hamiltonian and selected determinant space remain unchanged.

We evaluated the parallel performance of FLIP using communication-intensive FCI benchmarks on Fugaku.
No exposed communication waiting time was recorded on the slowest process in any of the 32 reported Hamiltonian--vector multiplications.
For N$_2$/cc-pVTZ with $2.62\times10^{12}$ determinants, Hamiltonian--vector multiplication achieved 91.3\% strong-scaling efficiency from 18,000 to 81,000 nodes, corresponding to more than 3.8 million compute cores at the largest allocation.
All four benchmark systems achieved endpoint strong-scaling efficiencies above 90\%.
The reductions in multiplication time were particularly pronounced in communication-intensive cases, demonstrating the effectiveness of combining local CI-coefficient access with nearby link retrieval.

We also obtained an FCI reference potential-energy curve for N$_2$/cc-pVTZ and demonstrated the structural compactness of the TPB representation.
At all six sampled bond lengths, selecting spin bitstrings according to their collective weights in reference HCI wavefunctions yielded variational TBSCI energies within 0.1~m$E_h$ of FCI while retaining fewer than 1\% of the FCI determinants.
The CI-coefficient distributions showed that expanding the selected spin-bitstring sets primarily increased the populations of the smaller-coefficient bins, whereas those of the largest-coefficient bins remained nearly unchanged.

Beyond the precomputed-link implementation demonstrated here, local construction of double-excitation links offers a route to substantially reducing communication volume and eliminating the need for persistent storage of precomputed double-link tables, paving the way for PT2 calculations involving external determinant spaces that can exceed the variational space by orders of magnitude.

\section*{Supplementary Material}
The Supplementary Material provides detailed timings and node-hour costs for both the original and FLIP algorithms in the four FCI strong-scaling benchmarks, covering Hamiltonian--vector multiplication (Tables~S1--S4) and Davidson execution (Tables~S5--S8).

\section*{Acknowledgments}
This work used computational resources of supercomputer Fugaku (Project ID: hp250291) and the Hokusai Big Waterfall2 system at RIKEN (Project IDs: RB230090 and RB250029).

\section*{Author declarations}
\subsection*{Competing interests}
The authors declare that they have no competing interests.

\section*{Data availability}
Executable binaries of the TBSCI eigensolver, together with
representative input and output data, are publicly available
at \url{https://github.com/enhua-xu/TBSCI-202611}.

\section*{Declaration of generative AI and AI-assisted technologies in the manuscript preparation process}

During the preparation of this work, the authors used ChatGPT
(OpenAI) to improve the language, clarity, and organization
of the manuscript. The authors reviewed and edited the content
and take full responsibility for the published article.

\bibliographystyle{elsarticle-num}
\bibliography{refs}

\end{document}